\documentclass[amsmath,amssymb,floatfix,twocolumn]{revtex4}
\usepackage{graphicx}
\usepackage{color}

\begin{document}

\title{Comparison of density functionals for nitrogen impurities in ZnO}

\author{Sung Sakong}
\email{sung.sakong@uni-due.de}
\author{Johann Gutjahr}
\author{Peter Kratzer}
\affiliation{Fakult\"at f\"ur Physik and  Center for Nanointegration (CENIDE), Universit\"at Duisburg-Essen, Lotharstra{\ss}e 1, 47057 Duisburg, Germany}

\begin{abstract}
Hybrid functionals  and empirical correction schemes are compared to conventional semi-local density functional theory (DFT) calculations in order to assess the predictive power of these methods concerning the formation energy and the charge transfer level of impurities in the wide-gap semiconductor ZnO.
While the generalized gradient approximation fails to describe the electronic structure  of the N impurity in ZnO correctly, 
methods that widen the band gap of ZnO by introducing additional non-local potentials yield the formation energy and charge transfer level of the impurity in reasonable agreement with hybrid functional calculations. 
Summarizing the results obtained with different methods, we corroborate earlier findings that the formation of substitutional N impurities at the oxygen site in ZnO from N atoms is most likely slightly endothermic under oxygen-rich preparation conditions, and introduces a deep level more than 1~eV above the valence band edge of ZnO.
Moreover, the comparison of methods elucidates subtle differences in the predicted electronic structure, e.g. concerning the orientation of unoccupied orbitals in the crystal field and the stability of the charged triplet state of the N impurity. Further experimental or theoretical analysis of these features could provide useful tests for validating the performance of DFT methods in their application to defects in wide-gap materials. 
\end{abstract}


\maketitle

\section{Introduction}
The electronic properties of impurities in semiconductors are a topic of great interest both in basic science and in technology. 
Ideally one would like to predict if and how a given impurity atom can be incorporated into a semiconductor, and whether it will be electronically active as a donor or acceptor. 
The answers to these questions are provided by the formation energy of the impurity (at least if sample preparation is close to thermodynamic equilibrium), and by the so-called charge transfer level, i.e., the position of the Fermi level in the electronic band gap for which two charge states of the impurity are equally probable  thermodynamically.
Hence, one would like to have a theoretical method at hand that allows one to accurately predict on the same footing the formation energy and  the spin multiplicity of charged impurity levels as well as the energetics of the charge transfer. 
First-principles methods using density functional theory (DFT) have brought us a big step further towards this goal, in particular when charge transfer levels are derived from total energy differences rather than from Kohn-Sham levels, but they still fall short of providing a complete and reliable description of the electronic properties. 
The problem is exacerbated in wide-gap semiconductors where conventional DFT methods often severely underestimate the band gap, and thus make the determination of charge transfer levels unreliable.
Here, we consider as a particular example the nitrogen impurity in ZnO, which has been studied thoroughly in experiment \cite{Tarun2011,Liu2011} and may serve as a reference system that illustrates the typical difficulties encountered.
Originally, N on the oxygen site (N$_{\mathrm{O}}$) was thought to be a promising candidate for achieving $p$-type doping in ZnO \cite{Look2002},
but there is growing evidence at present that, in fact, it acts as a deep level.
This change of view was promoted both by  experimental evidence \cite{Tarun2011} and by calculations using hybrid functionals \cite{Lyons2009,Gallino2010} and using Slater transition state method \cite{Furthmuller2012}, which yield a deep charge transfer level $\varepsilon(0/-)$, in contrast to more traditional DFT calculations within the generalized gradient approximation (GGA) that indicated a level 0.3--0.4~eV above the valence band \cite{Lyons2009,Furthmuller2012}. 
As shown in a comparative study by \'Agoston and Albe \cite{Agoston2009}, while hybrid functionals seem to remedy many of the problems of conventional DFT calculations of wide-gap semiconductors (in particular, they give values of the band gap close to the experimental ones), there are still open issues: 
One question concerns the fraction of exact exchange to be mixed in, taking into account that not only the position of the charge transfer level, but also the formation energy of the impurity from atomic or molecular sources should be reproduced correctly.
Moreover, the use of hybrid functionals is associated with considerable computational costs, in particular when a plane-wave basis set is used.
Hence, computationally cheaper alternatives are sought for. 
The oxygen-derived valence bands in ZnO are in GGA too high in energy due to erroneously strong hybridization with the Zn $3d$ levels.
The first idea coming to mind is a GGA+U correction~\cite{Adeagbo2010,Dudarev1998}, 
which pushes down the occupied Zn $3d$-levels, in agreement with recent photoemission studies \cite{Lim2012}. In addition to the GGA+U correction on the Zn $3d$ state, 
recently, a hole-state potential of $\lambda_\mathrm{hs}(1-n_{i,\sigma}/n_\mathrm{ref})$ is introduced 
to correct the O $2p$ and N $2p$ levels~\cite{Lany2010}. The reference occupation $n_\mathrm{ref}$ is determined by the occupation of the  $p$ orbitals of the of anion in the absence of holes.
The correction term $\lambda_\mathrm{hs}$ is selected in such a way as to be consistent with Koopman's theorem, i.e., one is trying to enforce a linear relationship between the energetic position and the filling factor of an atomic level after applying the Hubbard-U-type correction.  
Then, unoccupied $p$ levels, having an occupancy $n_{i,\sigma}$  smaller than $n_\mathrm{ref}$, are pushed upward. 
Yet another approach, also due to Lany, Raebiger and Zunger \cite{Lany2008NLEP}, aims at corrections of the band edges of ZnO by applying an additional non-local external potential (NLEP) to the Zn $4s$ and $3d$ as well as to the O $2p$ orbitals. In contrast to GGA+U and hole-state potential methods, the corrections applied in NLEP are independent of the occupation in the atomic sub-level, i.e. they apply equally to occupied and unoccupied states.  
In the present study, both the long-known GGA+U as well as the NLEP approach are discussed under the notion of `empirical correction schemes', since they allow the calculation of results in better agreement with experiment at computational costs comparable to plain GGA, at the price of introducing parameters that are material-specific, but can be determined empirically for each material under study.

\section{Computational details \label{sec:computational} } 

We perform spin-polarized DFT calculations within the GGA (PBE functional~\cite{Perdew1996}), the generalized Kohn-Sham scheme (HSE and PBE0 functionals~\cite{Heyd2003}), and various empirical correction schemes (GGA+U~\cite{Dudarev1998} and NLEP~\cite{Lany2008NLEP} methods) using the software package {\sc vasp}~\cite{Kresse1996, Blochl1994PRB, Paier2005, Paier2006}. A supercell of $3\times3\times2$ ZnO elementary unit cells  (Zn$_{36}$O$_{36}$) is selected for the host material, and a Monkhorst-Pack {\bf k}-point mesh of $2\times2\times2$ is used for the integration over the Brillouin zone of the supercell. This supercell, for which the computational demand is still manageable, also serves us as a benchmark to compare the performance  of different density functionals.
To test the dependence of the defect properties on the supercell size, we also use $4\times4\times3$ and $6\times6\times4$ ZnO elementary unit cells  (Zn$_{96}$O$_{96}$ and Zn$_{288}$O$_{288}$) for selected methods. 
For the GGA and the empirical correction methods, the wave functions are expanded in the plane wave basis set up to a kinetic energy of 400~eV. For hybrid functionals, the cut-off energy is reduced to 300~eV for speeding up the computations. Even with this reduced basis set, the computer time consumed by the hybrid functional calculations is several hundred times bigger  than the time used for the conventional GGA calculations. 

The calculations are carried out with a number of different methods, labelled M-1 to M-9 in the following.
The parameters applied for each calculation method are listed in Table~\ref{tab:method}. The GGA calculation (M-1) uses no additional parameters apart from those specified in the definition of the functional \cite{Perdew1996}. In addition to the widely used GGA-PBE, we have employed hybrid functionals with screened Fock exchange (M-2 and M-3) and without such screening (M-4) for a better description of the electronic band gap. 
In the screened hybrid calculations, two values for the mixing-in of the Fock term are chosen. The method M-2 uses the originally suggested amount of exact exchange (25\%)~\cite{Heyd2003}, and M-3 uses 36\%{}, bringing the band gap of pure ZnO into better agreement with experiment. As indicated by Burke \cite{Burke2012}, the M-3 can be also considered an empirical scheme, because of the material specific adjustment for a better agreement with the experimental band gap. 
While the lattice parameters of the supercell and the internal coordinates of the atoms have been optimized in the calculations in general, we have tested separately the effect of internal relaxations and of the supercell size in the case of the GGA+U scheme. This is of interest because many researchers choose to apply the GGA+U method as a correction scheme merely applied to the electronic structure on top of a geometry calculated by another scheme or adopted from experiment. The GGA+U calculations are performed with three parameter sets (M-5, M-6 and M-7). The M-5 and M-6 calculations commonly use the PBE lattice constant $a=a^\mathrm{PBE}$ with fixed atoms at the PBE configuration (M-5) or fully relaxed atoms (M-6). For M-7, we optimized both the lattice constant and the atomic positions. 
In the NLEP calculations, the parameters from Ref.~\onlinecite{Lany2008NLEP} are employed. 
Since NLEP applies a non-local correction to the O $2p$ orbitals, the question arises if such a correction should also be applied to the N $2p$ orbitals. We test both possibilities: 
In M-8, no corrections are made to the N atom potential, while in M-9 the same corrections as applied to the O atom are also made to N atom potential.

\begin{table*}[htdp]
\begin{center}
\begin{tabular}{ccccccccc}
\hline
method & functional & $a$ & coordinates & parameters\\
\hline
M-1 & PBE & optimized & relaxed & standard parameterization, see Ref.~\onlinecite{Perdew1996} \\
M-2 & HSE & optimized & relaxed &amount of exact exchange $\alpha=0.25$, cutoff radius $r=0.3$~\AA$^{-1}$\\
M-3 & HSE & optimized & relaxed &amount of exact exchange $\alpha=0.36$, cutoff radius $r=0.2$~\AA$^{-1}$ \\
M-4 & PBE0 & optimized & relaxed &amount of exact exchange $\alpha=0.25$, cutoff radius $r=0.0$~\AA$^{-1}$ \\
M-5 & GGA+U & $a^\mathrm{PBE}$ & fixed & $U-J=9.0$~eV on Zn $d$-orbital\\
M-6 & GGA+U & $a^\mathrm{PBE}$ & relaxed & $U-J=9.0$~eV on Zn $d$-orbital\\
M-7 & GGA+U & optimized & relaxed & $U-J=9.0$~eV on Zn $d$-orbital\\
M-8 & NLEP &optimized & relaxed & parameters from Ref.~\onlinecite{Lany2008NLEP} on Zn $spd$- and O $sp$-orbitals\\
M-9 & NLEP &optimized & relaxed &  M-8 + same parameters applied on N $sp$-orbitals as on O $sp$-orbitals \\
\hline
\end{tabular}
\end{center}
\caption{Details of the employed methods. $a^\mathrm{PBE}$ is the lattice constant optimized  with the PBE functional.}
\label{tab:method}
\end{table*}%

The formation energy of a charged N impurity in ZnO is evaluated using the expression 
\begin{equation}\label{eq:Ef}
E_\mathrm{f}[q,\varepsilon_\mathrm{F}]= E_\mathrm{tot}-E_\mathrm{host}-\sum_i \delta_i\mu_i+q(\varepsilon_\mathrm{F}+E_v) + E_\mathrm{corr.}
\end{equation}
where the Fermi level $\varepsilon_\mathrm{F}$ is considered to be a variable. $E_\mathrm{tot}$ und $E_\mathrm{host}$ represent the DFT total energy (relative to isolated atoms) with and without  the N impurity in the cell, respectively. The method of a homogeneous background charge is used to model charge states of the N impurity within the supercell approach. 
$\delta_i$ and $\mu_i$ are the change of the number of atoms in the cell and the  corresponding chemical potential for species $i$. 
In thermodynamic equilibrium with bulk ZnO, one of the chemical potentials can be eliminated using the relation $\mu_\mathrm{ZnO}=\mu_\mathrm{Zn}+\mu_\mathrm{O} \approx -E_\mathrm{ZnO}^\mathrm{coh} $ at zero temperature, where $E_\mathrm{ZnO}^\mathrm{coh}$ is the cohesive energy of ZnO per formula unit. 
The remaining variable $\mu_{\mathrm{O}}$ is fixed by the growth conditions: Under oxygen-rich processing, the chemical potential of oxygen is determined by the 
binding energy of the gas-phase oxygen molecule, $\mu_\mathrm{O}=-\frac{1}{2}E_\mathrm{O_2}$ and under oxygen-poor conditions, it is determined by the equilibrium of ZnO with  metallic zinc, $\mu_\mathrm{O}=-E_\mathrm{ZnO}^\mathrm{coh}+E_\mathrm{Zn}^\mathrm{coh}$. For the nitrogen chemical potential, we choose half the energy of an N$_2$ molecule or the energy of atomic N for N-poor and N-rich conditions, respectively. 
$E_v$ is the energy of the top of the valence band, as determined from the perfect, undoped ZnO supercell calculation.
Furthermore, the $E_\mathrm{corr.}=q \Delta V +\alpha_{Md} q^2 /2\epsilon L$ term \cite{Makov1995} is introduced to correct the spurious interaction between periodic images of charged impurities in the supercell approach, with the Madelung constant $\alpha_{Md}$, the dielectric constant $\epsilon$ and the distance to the next-neighboring periodic image $L$, as described in Ref.~\onlinecite{Sakong2011SST}. 

In the following sections, we need a common reference for energy levels obtained with different methods, e.g. when comparing charge transfer levels or the density of states (DOS). This requires some caution as discussed by Alkauskas, Broqvist and Pasquarello \cite{Alkauskas2008, Alkauskas2011}.  
While the energy zero for the Kohn-Sham levels may vary in the plane-wave method from one calculation to another, it is nonetheless possible to align the energy axis of different calculations (performed with the same functional) by comparing the electrostatic potential in the core region of equivalent atoms.
Moreover, 
by referring all energy levels to the vacuum level, it is even possible to plot the DOS of different methods on a common energy scale. 
With this aim, we determine for each method the position of the vacuum level from the calculated work function of a ZnO($10\bar{1}0$) slab consisting of 12 atomic layers. 
For this slab thickness, it is assured that the valence states of the inner layers are converged to the bulk valence state, making it possible to transfer the calculated position of the vacuum level to the corresponding 3D supercell with the impurity. 
For the calculations with the hybrid functionals and the GGA+U method (except M-7), the fixed atomic configuration of the PBE calculation is used for constructing the slab with the corresponding lattice constants. In case of the GGA+U methods, the reason for working with a fixed geometry is the observation that adding the +U term strongly affects the lattice parameters, and in case of a slab allowing for atomic relaxation would lead to an unnatural contraction of the slab along the surface normal. To avoid such artifacts we neglect the relaxation in the calculations M-5 and M-6. In case of the hybrid functional calculations, the purpose of neglecting relaxations in the slab was simply to save computation time.  
For the other methods, the slab is fully relaxed with the corresponding lattice parameters.

\section{Results}
\subsection{Formation energies}
\begin{figure}[htbp]
\begin{center}
\includegraphics[width=7.5cm]{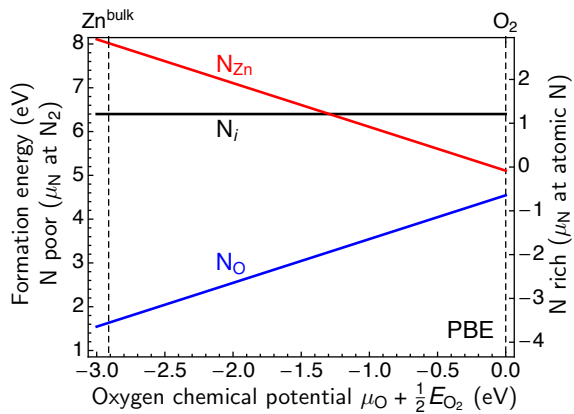}
\caption{Formation energy of N impurities in ZnO calculated using the  PBE functional as function of the oxygen chemical potential. 
}
\label{fig:eform}
\end{center}
\end{figure}

We start with the determination of the most stable lattice site at which the nitrogen impurity will be incorparated in ZnO. To this end, we  calculate and compare the formation energy of the neutral N impurity at the O lattice site (N$_\mathrm{O}$), at the Zn lattice site (N$_\mathrm{Zn}$) and at the interstitial site (N$_i$) using the method M-1 and the Zn$_{36}$O$_{36}$ cell. 
Figure~\ref{fig:eform} shows the formation energies of N$_\mathrm{O}$, N$_\mathrm{Zn}$ and N$_i$ as a function of the oxygen chemical potential $\mu_\mathrm{O}$. 
We observe that the N$_\mathrm{O}$ impurity is the most stable species under the usual growth conditions of high oxygen pressure. 
This is different from the carbon impurity in ZnO, where C$_{\mathrm{Zn}}$ was found to be more stable than C$_{\mathrm{O}}$ under almost all conditions~\cite{Sakong2011SST}. In chemical language, this finding indicates that the carbonate (CO$_3^{2-}$) structure formed in case of C$_{\mathrm{Zn}}$ is much more efficient in lowering 
the formation energy of C$_{\mathrm{Zn}}$ than the nitrate-like NO$_3^-$ structure formed in N$_{\mathrm{Zn}}$. 
Moreover, we find that the N$_i$ impurity has a larger formation energy  than  N$_{\mathrm{O}}$ 
over the whole $\mu_{\mathrm{O}}$ range. 
The relative stability of the different impurity types remains the same when the hybrid functional M-2 is used. This knowledge about the most stable incorporation site provides the justification for the  experimentalists'  assignment of their observed deep charge transfer level $\varepsilon(0/-)$ to the N$_{\mathrm{O}}$ impurity. 
Therefore, we focus on the N$_{\mathrm{O}}$ impurity for the comparison of the various functionals and proposed correction schemes.

The properties of the ZnO host material and the N$_{\mathrm{O}}$ impurity obtained with various methods 
are listed in Table~\ref{tab:Eform}.  The  neutral N$_{\mathrm{O}}$ impurity is spin-polarized with a magnetic moment of 1~$\mu_{\mathrm{B}}$. 
Most notably, it is found that the substitution of oxygen by a nitrogen atom from an atomic N source is exothermic in the PBE and GGA+U method, while it is calculated to be endothermic when using hybrid functionals or NLEP. 
The differences between the GGA+U schemes with and without geometry optimization are relatively small compared to the overall variation of the results. Inclusion of internal relaxations has very little effect (see M-5 and M-6). The optimization of the supercell volume in GGA+U (M-7) worsens the agreement of the lattice parameter with experiment, but leads to a larger, and hence more realistic, band gap. 
Using an increased amount of exact exchange (M-3) results in a strong increase of the formation energy, i.e., it makes the defect formation even more endothermic. The issue of screening the Fock exchange seems to be less important, since the formation energy calculated with M-4 is just $\sim$0.2~eV larger than the M-2 value when the same mixing parameter $\alpha=0.25$ is used. 
When the N$_2$ molecule is used as reference for $\mu_{\mathrm {N}}$, a somewhat different picture evolves.
In the N-poor limit
the N$_\mathrm{O}$ impurity formation is always strongly endothermic, i.e., the formation energy is positive (column 9  in Table~\ref{tab:Eform}). 
Compared to the formation energy of 4.55~eV in PBE (M-1), the formation energy of the neutral N$_\mathrm{O}$ impurity is around one eV larger in hybrid functionals and in NLEP. In the GGA+U method, a modest increase of 0.6~eV with respect to PBE is noticed. 
The variation of the formation energy among the different methods is less drastic in the N-poor case as compared to the N-rich case.

\begin{table*}[htdp]
\begin{center}
\begin{tabular}{ll|ccl|crccl|ccc}
\hline
 & &V$_\mathrm{cell}$&$E_g$ &$\epsilon$ & $\Delta E$ &\multicolumn{2}{c}{  N-rich (eV)} &\multicolumn{2}{c}{  N-poor (eV)} & \multicolumn{2}{|c}{$\varepsilon(0/-)$ (eV)} & $\varepsilon(0/-)/E_g$ \\
method & functional & (\AA$^{3}$)& (eV) & & &$E_\mathrm{f}[0,\varepsilon_\mathrm{F}]$ & $\Delta \mu$&$E_\mathrm{f}[0,\varepsilon_\mathrm{F}]$ & $\Delta \mu$& w/ $E_\mathrm{corr.}$ & w/o $E_\mathrm{corr.}$\\
\hline
M-1  & PBE  &48.27 &0.77 &10.28 &  2.39 & $-0.64$ & 3.04 & 4.55 & $-2.15$& 0.46 &0.34 & 0.60 \\
M-2  & HSE & 47.02& 2.29 & $7.21^{a}$ & 3.34 & 0.69  & 2.65& 5.42  & $-2.08$ & 0.97 & 0.83& 0.42 \\
M-3 & HSE &46.53& 3.37& -- & 5.38 & 2.73 & 2.65 & 5.90 & $-0.52$& 1.46 & 1.32 & 0.43 \\
M-4   & PBE0& 46.86& 3.18 & -- & 3.52 & 0.88 & 2.64 &5.60  & $-2.07$& 1.23 &$\varepsilon(+/-)$ & 0.42 \\
M-5  &GGA+U &48.27 &1.78 &10.73 & 2.95 & $-0.09$  & 3.04& 5.10  & $-2.15$& 0.72 & 0.62 & 0.40 \\
M-6 &GGA+U&48.27 &1.78 &10.73 & 2.92 & $-0.12$  & 3.04& 5.07  & $-2.15$& 0.73 & 0.63 & 0.41  \\
M-7  &GGA+U&42.86 &2.20 &8.72  & 3.00 & $-0.03$ & 3.04 & 5.16 & $-2.15$& 0.80 & 0.67  & 0.36 \\
M-8 &NLEP&43.49&3.30 &6.09 & 4.46 & 0.54 & 3.92 & 5.73  & $-1.27$& 1.13 & 0.95 & 0.34 \\
M-9  &NLEP &43.49&3.30 &6.09 & 4.10 & 0.18  & 3.92 & 6.89  & $-2.79$& 1.12 & 0.93 & 0.34 \\
Exp.    & & 47.61& 3.44 & $8.91^{b}$& n.a. & n.a. & 2.58$^{c}$ & n.a.  & $-2.31^{c}$  
& \multicolumn{2}{c}{1.1 -- 1.3$^{d}$} & 0.35 \\
\hline
{\scriptsize $^a$Ref.~\onlinecite{Paier2008} } \\
{\scriptsize $^b$Ref.~\onlinecite{Ashkenov2003} }\\
{\scriptsize $^c$Ref.~\onlinecite{CRCHandbook} }\\
{\scriptsize $^d$Ref.~\onlinecite{Tarun2011,Garces2002} }
\end{tabular}
\end{center}
\caption{Crystal properties (unit cell volume V$_\mathrm{cell}$, band gap $E_g$, and dielectric constant $\epsilon$) of ZnO, formation energy of the neutral N$_\mathrm{O}$ impurity, and charge $\varepsilon(0/-)$ level using various methods. The formation energy is calculated under oxygen-rich conditions, as described in the text. The details of the methods are found in Table~\ref{tab:method}. The last row contains experimental data; n.a. stands for `not available'.
}
\label{tab:Eform}
\end{table*}

This raises the question whether the variation of the formation energy among the different methods is due to a different description of bonding in the solid, or merely due to a different description of bonding in the molecules used to define the chemical potentials. 
In order to trace the causes for the differences in formation energy, we decompose it into two contributions, 
$E_\mathrm{f} = \Delta E - \Delta \mu$, motivated by the three leading terms in Eq.~\ref{eq:Ef}.
(Note that the last two terms in Eq.~\ref{eq:Ef} vanish for a {\em neutral} impurity).  
The first contribution, $\Delta E=E_\mathrm{tot}-E_\mathrm{N}^{\rm atom}-(E_\mathrm{host}-E_\mathrm{O}^{\rm atom})$, arises due to the difference in bond strength between the O and the N atom to its neighbors in the crystal. The second one, $\Delta \mu=\mu_\mathrm{N}-E_\mathrm{N}^{\rm atom}-(\mu_\mathrm{O}-E_\mathrm{O}^{\rm atom})$, is due to the different levels of description of the molecular O$_2$ bond that appears in the thermodynamic reservoir term in the N-rich limit. The energies of atomic O and N ($E_\mathrm{O}^{\rm atom}$ and $E_\mathrm{N}^{\rm atom}$) are introduced to define a common reference in both the crystal and the molecules. In the formation energy $E_\mathrm{f}$, the atomic energies appearing in $\Delta E$ and $\Delta \mu$ cancel out. 

As can be seen from Table~\ref{tab:Eform}, $\Delta E$ is of the order of 2.9~eV to 3.5~eV for most of the methods studied (M-2, M-4, M-5, M-6, M-6). 
The PBE functional (M-1) yields the smallest $\Delta E$, while $\Delta E$  is largest in the methods M-3 (HSE), M-8 and M-9 (NLEP).  
We note that $\Delta E$, which is independent of thermodynamic reservoirs and can be seen as a descriptor for the difference in bond strengths of O and N atoms to Zn atoms, follows the same trend as the value of the band gap $E_g$.  Thus, the methods that underestimate the band gap also tend to underestimate differences in chemical bonding. 

Next, we consider $\Delta \mu$. Since this quantity reflects the bond strength in the O$_2$ molecule (N-rich limit), or the difference in bond strength of the O$_2$ and the N$_2$ molecule (N-poor limit), it can be compared to experimental data.  
In the N-rich case, 
the hybrid functionals (M-2, M-3 and M-4) give the most accurate estimate of the bond strength. The PBE (M-1), and likewise the empirical correction schemes, tend to overestimate the bond strength of O$_2$ which leads to a less accurate description of the thermodynamic reservoir. 
Also in the N-poor case, the hybrid functionals M-2 and M-4 provide good estimates of $\Delta \mu$, i.e. both the O$_2$ and N$_2$ bond strengths are well described. The  PBE (and, identically, GGA+U) results come closest to the experimental value of $\Delta \mu$, because the overestimated O$_2$ bond strength is compensated by the likewise overestimated N$_2$ bond strength. 
We note also that mixing in a larger amount of exact exchange (M-3) results in a drastically underestimated N$_2$ bond strength. 
The NLEP method results in large deviations of $\Delta \mu$ from its experimental value, but the absolute magnitude of the deviation is somewhat smaller after corrections to the energetic position not only of the O orbitals (M-8), but also of the N orbitals (M-9) have been introduced. 

Combining the results for $\Delta E$ and $\Delta \mu$, we conclude that the smaller variation of the formation energy among the different methods in the N-poor limit
is due to a partial compensation between the $\Delta E$ term and the $\Delta \mu$ term entering into the formation energy when the molecular N$_2$ bond is taken into account: In M-3, for example, both Zn-N and N$_2$ bonds are estimated to be weaker than in the other methods. 
We note that only the hybrid calculations of M-2 and M-4 predict the reference energy of the thermodynamic reservoirs within 0.2~eV of the experimental values, and these methods yield a stronger Zn--O bond compared to Zn--N than conventional GGA. We therefore suppose that M-2 and M-4 give the most reliable estimates of $E_\mathrm{f}$ in both the N-rich and the N-poor limit. The NLEP method (M-8) comes close to these results, but including an additional correction to the energy of the N orbitals (M-9) apparently does not give an improved description of $E_\mathrm{f}$.

\subsection{Orbital structure of the N$_\mathrm{O}$ impurity}

\begin{figure}[thbp]
\begin{center}
\includegraphics[width=8.5cm]{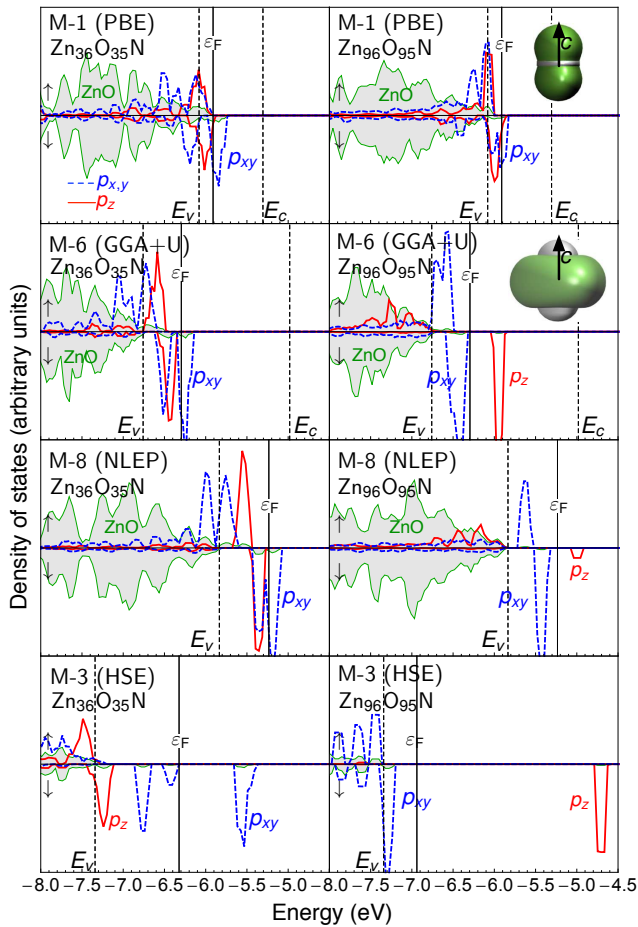}
\caption{Orbital-projected density of states of a neutral N$_{\mathrm{O}}$ impurity in ZnO, calculated in the  Zn$_{36}$O$_{35}$N supercell (left column) and in the Zn$_{96}$O$_{95}$N supercell (right column). Vertical dashed and solid lines indicate the band gap of the host material and the Fermi level $\varepsilon_{\mathrm{F}}$ of the system, respectively.  The energy is expressed with respect to the corresponding vacuum level in each method.
 The charge densities of the N impurity in the $4\times4\times3$ supercell (Zn$_{96}$O$_{95}$N) are presented using the M-1 (PBE) and the M-6 (+U) methods. The green and white isosurfaces represent the charges of occupied and unoccupied states at $0.3e$, respectively. The charge of occupied states is integrated from $E_v$ to $\varepsilon_\mathrm{F}$ and the charge of unoccupied states is integrated from $\varepsilon_\mathrm{F}$ to $E_c$. The arrows show the $c$-direction of the cell.
}
\label{fig:dos}
\end{center}
\end{figure}

\begin{figure}[htbp]
\begin{center}
\includegraphics[width=8.5cm]{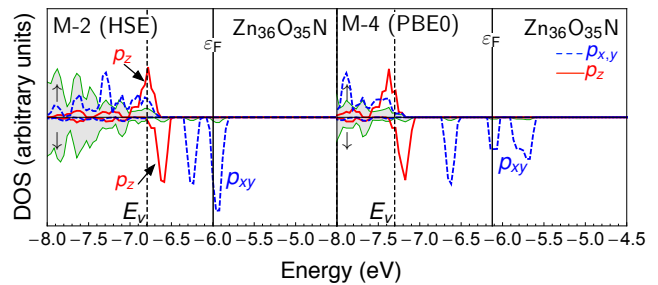}
\caption{Orbital-projected density of states of the neutral N$_{\mathrm{O}}$ impurity calculated in the ZnO (Zn$_{36}$O$_{35}$N) supercell using hybrid density functionals.
}
\label{fig:dosHSE}
\end{center}
\end{figure}

Before discussing the charged states of the N$_\mathrm{O}$ impurity, we first analyze the orbital character of the defect states introduced in the band gap of ZnO.
The density of states (DOS) plots for the neutral N$_\mathrm{O}$ impurity obtained from the  various methods are presented in Figs.~\ref{fig:dos} and \ref{fig:dosHSE}. 
Since nitrogen has one electron less than oxygen, the N$_{\rm O}$ impurity in ZnO misses one electron in the 2$p$ shell.
In the wurtzite crystal symmetry, the degeneracy of the $2p$ orbitals of N is split into $2p_z$ (along the $c$-direction) and $2p_{x,y}$ (in the  $ab$-plane). The $2p_{x,y}$ orbitals of N will hybridize with the $2s$ orbital to form $sp^2$ hybrids pointing toward the nearest Zn neighbors in the hexagonal $ab$-plane, whereas the $2p_z$ orbital is oriented toward the nearest Zn atom along the $c$-axis. 
The missing electron leads to one unoccupied  N orbital, either $2p_{x,y}$ or $2p_z$. Differences in the orbital occupancies among the methods arise due to the different energetic positions of the defect orbitals with respect to the host valence band, as well as due to the different magnitude of the exchange splitting.
As discussed by Lany and Zunger \cite{Lany2010}, a  Zn$_{96}$O$_{96}$ (or larger) supercell  is needed to obtain the correct orbital occupancy in agreement with the electron paramagnetic resonance (EPR) characterization of N-doped ZnO \cite{Carlos2001}.
The pertinent results are shown in the right column of Fig.~\ref{fig:dos}. 
In the GGA+U (M-6), NLEP (M-8) and HSE (M-3) calculations, the unoccupied orbital points along the $c$-axis (cf. the isosurfaces in Fig.~\ref{fig:dos}), in agreement with experiment. This is also reflected in the Jahn-Teller splitting, which leads to a longer N--Zn bond in $c$-direction than in the $ab$-plane (see Table~\ref{tab:cellsize}, rows 6 and 7). 
Our results agree well with the B3LYP hybrid functional calculations by Gallino {\it et al.}~\cite{Gallino2010, Gallino2010a} who used large 108 and 192 atom supercells. We expect the methods M-3 and M-4 to show the same behavior. 
However, in the PBE functional (M-1) calculations, as shown in Fig.~\ref{fig:dos} and Table~\ref{tab:cellsize}, no Jahn-Teller splitting is obtained, and the $2p_{x,y}$ and $2p_z$ impurity orbitals come to lie in energetically overlapping regions, due to the overestimation of their hybridization with the valence band. 
Thus, PBE fails to correctly describe the orbital structure of the N impurity as shown in Fig.~\ref{fig:dos}.

In order to reduce the computational cost, one would like to work with an as small as possible supercell. 
In the Zn$_{36}$O$_{36}$ cell, the formation energy of the neutral N$_{\rm O}$ impurity is already converged to better than 0.1~eV, as can be seen from Table~\ref{tab:cellsize}. However, in this smaller cell, the occupancy of the impurity orbitals disagrees with the experimental findings for all functionals used in our study: 
The electron is missing in the $ab$-plane, while the $2p_z$ orbital is fully occupied.
In this small cell, the spin-minority $sp^2$ orbitals form impurity bands due to the interaction with their periodic images (double-peak feature labelled $p_{x,y}$ in the plots in Fig.~\ref{fig:dos}, left column).  
The orbital occupancy also affects the magnitude of the exchange splitting: 
When the $2p_z$ orbital is unoccupied, i.e. in the calculations using the Zn$_{96}$O$_{96}$ cell, a large exchange splitting between the spatially well-localized $2p_z$ orbitals of the majority and minority spin electrons is observed. However, in the smaller  Zn$_{36}$O$_{36}$ cell, the partially occupied $p_{x,y}$ orbitals experience only a smaller exchange splitting, because they are part of extended bands in the $ab$-plane. 
The comparison of the different methods (see Figs.~\ref{fig:dos} and \ref{fig:dosHSE}) reveals that the hybrid functional calculations (M-2, M-3, and M-4) generally lead to  a stronger exchange splitting than the PBE functional (M-1) and the empirical correction schemes (M-6 and M-8). 

\subsection{Charge transfer levels}

In the following, we discuss the charge states of the N$_{\mathrm{O}}$ impurity.
The charge transfer levels are calculated from the total energies, i.e., by determining the value of the Fermi energy $\varepsilon_\mathrm{F}$ for which the formation energies of two different charge states in Eq.~\ref{eq:Ef} become equal. In particular, the position of the acceptor level is determined by the formation energies of the neutral and the negatively charged impurity. 
First, we consider the neutral impurity. In this case, the correction term $E_\mathrm{corr.}$ is zero, as discussed above. 
As listed in Table~\ref{tab:cellsize}, the formation energy of the neutral impurity changes by less than 10~meV when going from the Zn$_{96}$O$_{96}$ cell  to the Zn$_{288}$O$_{288}$ cell in case of the GGA+U (M-6) and NLEP (M-8) calculations. Thus, it is well converged. In the PBE (M-1), the energy converges somewhat slower with the cell size than in the other two methods. 
For the negatively charged impurity, even  the large Zn$_{288}$O$_\mathrm{288}$ supercell is not sufficient to reach the same level of convergence of the formation energy as for the neutral impurity. 
This is due to the spurious long-range interaction of the charged impurity with its periodic images. In order to improve the convergence with cell size, we employ a correction term which is meant to compensate for the Madelung energy of a lattice of point charges in a dielectric medium \cite{Makov1995}, as was used already in Ref.~\onlinecite{Sakong2011SST}. 
As seen from Table~\ref{tab:Eform}, applying such a correction term $E_{\rm corr.}$ places the acceptor level $\varepsilon(0/-)$ around 0.1~eV  deeper in the gap in all methods. We note that the magnitude of the correction correlates with the band gap $E_g$. This is because the correction term comprises the method-specific dielectric constant $\epsilon$, which is inversely correlated with the magnitude of the band gap.
Interestingly, in the PBE0 (M-4) calculation, the neutral state would be unstable if the $E_\mathrm{corr.}$ term were neglected; and a direct change from the positive to the negative state would be observed (denoted as $+/-$ in Table~\ref{tab:Eform}). 
The convergence of the charge transfer levels with the cell size is summarized in Table~\ref{tab:cellsize}. 
In the GGA+U (M-6) and NLEP (M-8) method, the value of $\varepsilon(0/-)$ obtained from the small Zn$_{36}$O$_\mathrm{36}$ cell is found to be in better agreement with the value obtained from the Zn$_{288}$O$_\mathrm{288}$ cell after applying the correction. Thus, the correction term is useful as it enables us to estimate the acceptor level already from a relatively small cell. For the PBE method (M-1), for which the defect states are unrealistically extended in space, the correction methods appears to be inapplicable.  

\begin{table*}[htdp]
\begin{center}
\begin{tabular}{lc|rrr|rrr|rrr}
\hline
Zn$_{n}$O$_{n-1}$N& & \multicolumn{3}{c}{PBE (M-1)} &\multicolumn{3}{|c}{GGA+U (M-6)} & \multicolumn{3}{|c}{NLEP (M-8)}\\
number of Zn-O units $n$  & & 36 &96 & 288 &36 &96 & 288 &36 &96 & 288  \\
\hline
$E_\mathrm{f}[q=0,\varepsilon_\mathrm{F}=E_v]$ & (eV) & $-0.64$ & $-0.59$ & $-0.57$ & $-0.12$ & $-0.15$ & $-0.16$ & 0.54 & 0.48 & 0.48\\
$E_\mathrm{f}[q=-1,\varepsilon_\mathrm{F}=E_v]$ without $E_\mathrm{corr.}$ & (eV) & $-0.30$ & $-0.27$ & $-0.25$ & 0.51 & 0.55 & 0.58 & 1.49 & 1.54 & 1.64\\
$E_\mathrm{f}[q=-1,\varepsilon_\mathrm{F}=E_v]$ with $E_\mathrm{corr.}$ & (eV) & $-0.19$ & $-0.15$ & $-0.19$ & 0.61 & 0.59 & 0.63 & 1.67 & 1.64 & 1.73\\
$\varepsilon(0/-)$ without $E_\mathrm{corr.}$  & (eV) & 0.34 & 0.32 & 0.32 & 0.63  & 0.69 &0.73& 0.95 & 1.07 & 1.15 \\
$\varepsilon(0/-)$ with $E_\mathrm{corr.}$     & (eV) & 0.46 & 0.44 & 0.38 &  0.73 & 0.74 & 0.78 & 1.13 & 1.16 & 1.25\\
$d_\mathrm{N-Zn}$ in $ab$-plane & (\AA) & 1.95 & 1.95 & 1.95 & 1.98 & 1.96 & 1.96 & 1.97 & 1.96 & 1.97\\
$d_\mathrm{N-Zn}$ in $c$-axis       & (\AA) & 1.94 & 1.94 & 1.94 & 1.95 & 2.16 & 2.16 &1.97 & 2.01 & 2.01 \\
\hline
\end{tabular}
\end{center}
\caption{The properties of the N$_\mathrm{O}$ impurity in Zn$_{36}$O$_{36}$, Zn$_{96}$O$_{96}$, and Zn$_{288}$O$_{288}$ supercells.}
\label{tab:cellsize}
\end{table*}%

\begin{figure}[htbp]
\begin{center}
\includegraphics[width=8.cm]{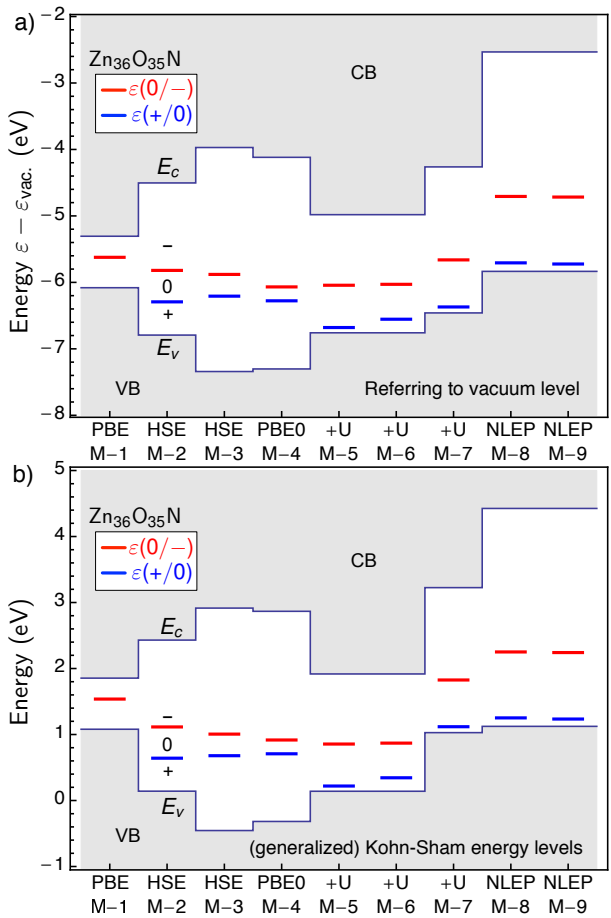}
\caption{$\varepsilon(+/0)$ and $\varepsilon(0/-)$ charge transfer levels of the N$_\mathrm{O}$ impurity in ZnO. The band edges are given with respect to  the vacuum level in a), and are  taken directly the from (generalised) Kohn-Sham levels in b). }
\label{fig:charged}
\end{center}
\end{figure}

For a systematic overview of the charge transfer levels as obtained from the different methods, we employ the  Zn$_{36}$O$_{36}$ cell.  
In Fig.~\ref{fig:charged}, the $\varepsilon(0/-)$ and $\varepsilon(+/0)$ levels obtained from the various functionals and correction schemes are displayed. 
The values of $\varepsilon(0/-)$ listed in Table~\ref{tab:Eform} quote the charge transfer level with respect to the valence band edge $E_v$, as obtained within each method. 
Thus, one needs a consistent scheme to compare energy levels obtained from different methods. Two approaches are mostly used in the literature: 
In Fig.~\ref{fig:charged}a, the vacuum levels as obtained from each method are aligned, as described in Section~\ref{sec:computational}.
As a simpler alternative, in Fig.~\ref{fig:charged}b the band edges in each method are taken directly from the  calculated  (generalized, in case of M-2, M-3, and M-4) Kohn-Sham energy levels without any further alignment.  
Using the first, more elaborate approach, we notice that the charge transfer levels $\varepsilon(0/-)$ in methods M-1 to M-7 are not as different as one might have expected; they all fall into an energy interval of 0.45~eV \cite{Alkauskas2011}.
This energy interval is somewhat larger (0.68~eV) if the simpler approach 
is used.
The NLEP method (M-8 and M-9) must be considered separately, as it attempts to correct the band gap only, but not the positions of the valence band edge.
We note that the energetic position of the valence band (VB) and conduction band (CB) edges display much stronger variations from one method to the other than the charge transfer levels. 
This is because the charge transfer levels have been determined from DFT total energies that are less susceptible to the self-interaction error in DFT, whereas the band edges are taken from (possibly re-aligned) Kohn-Sham eigenvalues. Different from the total energy, the latter are not variational with respect to the charge density, and are sensitive to the specific division between kinetic and potential energy of the electrons inherent to  each method. 

While the absolute value of the acceptor level varies considerably from method to method, the relative position within the band gap is rather similar (cf. Table~\ref{tab:Eform}). 
Due to the difficulty in determining band edges in DFT, the  $\varepsilon(0/-)$ charge transfer level erroneously appears as a level close to the valence band edge in PBE (M-1), only about 0.4~eV above the VB edge, whereas all other methods indicate $\varepsilon(0/-)$ to be a deep level.  
The highest absolute value of 1.46~eV is obtained using HSE with a value of $\alpha = 0.36$ (M-3), in agreement with a previous HSE calculation~\cite{Lyons2009}. 
From the other methods tested, only PBE0 (M-4) and NLEP (M-8 and M-9) arrive at a value larger than 1~eV. The variants of the GGA+U method (M-5, M-6 and M-7) give the correct trend, i.e. a deeper charge transfer level than in plain GGA, but in these methods the VB edge is still too high, and thus the charge transfer level appears too close to the VB edge. 
For comparison with experiment, the measured threshold for photoluminescence excitation of 2.4~eV~\cite{Tarun2011} or 2.22~eV~\cite{Garces2002} should be used to determine the position of $\varepsilon(0/-)$ relative to the CB edge. Allowing for a correction of $\sim 0.3$~eV for relaxation effects (see Ref.~\onlinecite{Lyons2009}), one arrives at an estimate of 1.1 -- 1.3~eV for the position of $\varepsilon(0/-)$ relative to the VB edge. 
In our judgement, the best agreement with the experimental results is obtained with the PBE0 functional (M-4), with the computationally much  more affordable NLEP method (M-8) yielding results of comparable quality.

It is interesting to observe that, apart from PBE, all methods also find the $\varepsilon(+/0)$ charge transfer level in addition to $\varepsilon(0/-)$. 
Moreover, in the hybrid functionals (M-2, M-3 and M-4), the energy range where the neutral N$_{\mathrm{O}}$ is stable turns out to be very limited. This range is smallest in PBE0 (M-4), i.e., for an admixture of unscreened Fock exchange, and becomes larger in the HSE functionals where Fock exchange is screened, growing with decreasing value of the mixing parameter $\alpha$. In contrast, the GGA+U and the NLEP methods place the $\varepsilon(+/0)$ charge transfer level close to the VB edge. 
The positively charged N$_{\mathrm{O}}$ impurity, which lacks two electrons compared to the closed-shell  O$^{2-}$ ion, 
is found to be a spin triplet, in accordance with Hund's rule, in all these methods. We note that a triplet state has been observed in N-doped samples e.g. in the 
green emission at 2.5 eV in optically detected magnetic resonance \cite{Carlos2001}. 

From a practical perspective, 
the energetic position of the $\varepsilon(+/0)$ charge transfer level is important for assessing the role of N$_{\mathrm{O}}$ impurities in samples where $p$-type doping has been achieved in some other way  (e.g., by $N_{\rm Zn} V_{\rm O}$ complexes, see Ref.~\onlinecite{Liu2012}): If the Fermi energy is brought below the $\varepsilon(+/0)$ level, this level will act as a `hole trap', i.e. it will reduce the number of holes in the VB that can be delivered by ionization of any shallow acceptors. 
Thus, in particular if the hybrid functionals predicted the $\varepsilon(+/0)$ correctly, N$_{\rm O}$ not only would be inoperative as an acceptor, but would even compensate other acceptors to be introduced into ZnO. 
However, we consider it more likely that the $\varepsilon(+/0)$ level is actually located rather close to the VB edge, as a recent experiment 
detected several candidates for donor-like hole traps below $E_v + 0.5$eV by optically excited minority carrier transient spectroscopy~\cite{Muret2012}. The admixture of exact exchange in hybrid functionals may well overestimate the stability of high-spin states such as the N$_{\rm O}^+$ triplet in comparison to experiment.

\section{Conclusions}

We have presented a comparative study of the performance of density functionals applied to the N impurity in ZnO. 
The standard GGA functional has difficulties in reproducing the experimental findings of a deep $\varepsilon(0/-)$ charge transfer level and of the $c$-orientation of the neutral N impurity orbital, because of the severely underestimated band gap. 
For a good estimate of both the formation energy and the acceptor level of N$_{\rm O}$, hybrid functionals are the best choice. In particular, PBE0 performs well in all respects, while using HSE with a large ($>25$\%) amount of exact exchange raises the formation energy to values that appear too large.
Exploring the computationally inexpensive empirical correction schemes GGA+U and NLEP, we noticed  that these schemes, in particular NLEP, already improve many properties of the N impurity in the ZnO semiconductor,  
e.g. they yield a larger band gap, a deeper $\varepsilon(0/-)$ charge transfer level, and the correct orientation of the neutral N impurity orbital. 
However, there are also notable differences between the empirical schemes and the hybrid functional calculations,  e.g. the absolute positions of energy levels with respect to vacuum. 
Moreover, the N$_{\rm O}^+$ spin triplet is predicted  much deeper inside the band gap by the hybrid functionals. Experimental identification of this level could provide an interesting additional test for our present understanding of the electronic structure of defects in wide-gap semiconductors. 

\section*{Acknowledgments}

The authors thank Stephan Lany for the NLEP code and Voicu Popescu for valuable discussions. 
P.K. is grateful to the Institute of Pure and Applied Mathematics (IPAM) for supporting his participation in the program 'Materials Defects: Mathematics, Computation and Engineering'.
The authors acknowledge the Center for Computational Sciences and Simulation (CCSS) of University Duisburg-Essen for computer time and the Deutsche Forschungsgemeinschaft DFG for financial support. 


\end{document}